\documentclass{emulateapj}

\usepackage{psfig,natbib}

\usepackage{apjfonts}

\newcommand{\kms}       {km~s$^{-1}$}
\newcommand{\h}        {$h^{-1}_{70}\,$~kpc}

\newcommand{\apg}       {^{>}_{\sim}}

\newcommand{\qsoabg} {SDSS~J095454.99+373419.9}   
\newcommand{\qsoafg} {SDSS~J095454.73+373419.7}   

\newcommand{\qsobbg} {SDSS~J083649.55+484154.0}   
\newcommand{\qsobfg} {SDSS~J083649.45+484150.0}   

\newcommand{\qsocbg} {SDSS~J231253.03+144453.4}   
\newcommand{\qsocfg} {SDSS~J231252.79+144458.6}   

\newcommand{\qsodbg} {SDSS~J211230.33$-$063332.1} 
\newcommand{\qsodfg} {SDSS~J211229.31$-$063331.4} 

\newcommand{\wAssoc} {$W_r$(assoc)}

\shorttitle{ }
\shortauthors{D. V. Bowen et al}

\received{March 24, 2006}
\accepted{May 30, 2006}

\begin{document}

\title{QSO Absorption lines from QSOs\footnotemark[1]}

\author{David V.~Bowen\altaffilmark{1}, 
Joseph F.~Hennawi\altaffilmark{1,2,3}, 
Brice M\'{e}nard\altaffilmark{4}, 
Doron Chelouche\altaffilmark{4,5}, 
Naohisa Inada\altaffilmark{6},
Masamune Oguri\altaffilmark{1},
Gordon T.~Richards\altaffilmark{7},
Michael A.~Strauss\altaffilmark{1},
Daniel E.~Vanden Berk\altaffilmark{8},
Donald G.~York\altaffilmark{9}}


\footnotetext[1]{Based in part on observations obtained with the Apache
Point Observatory 3.5-meter telescope, which is owned and operated by
the Astrophysical Research Consortium. }

\altaffiltext{1}{Princeton University Observatory, Princeton, NJ 08544} 

\altaffiltext{2}{Hubble Fellow}

\altaffiltext{3}{Dept.\ of Astronomy, University of California, Berkeley,
  CA 94720}

\altaffiltext{4}{Institute for Advanced Study, Einstein Drive, Princeton, NJ 08540}

\altaffiltext{5}{Chandra Fellow} 

\altaffiltext{6}{Institute of Astronomy, Faculty of Science, 
University of Tokyo, 2-21-1 Osawa, Mitaka, Tokyo 181-0015, Japan}

\altaffiltext{7}{Dept.\ of Physics and Astronomy, Johns Hopkins University, 3400
  N.~Charles St., Baltimore MD~21218}

\altaffiltext{8}{Department of Astronomy and Astrophysics, The Pennsylvania
  State University, 525 Davey Laboratory, University Park, PA~16802}

\altaffiltext{9}{Dept.\ of Astronomy and Astrophysics, University of
  Chicago,  Enrico Fermi Institute, 5640 South Ellis Avenue, Chicago, IL~60637.} 

\begin{abstract}
  
  We present the results of a search for metal absorption lines in the spectra
  of background QSOs whose sightlines pass close to foreground QSOs. We detect
  Mg~II~$\lambda\lambda 2796,2803$ absorption in {\it Sloan Digital Sky
    Survey} (SDSS) spectra of four $z\:>\:1.5$ QSOs whose lines of sight pass
  within 26$-$98~\h\ of lower redshift ($z\:\simeq\: 0.5-1.5$) QSOs.  The
  100\%\ [4/4 pairs] detection of Mg~II in the background QSOs is clearly at
  odds with the incidence of associated $z_{\rm{abs}}\:\simeq\:z_{\rm{em}}$
  systems --- absorbers which exist towards only a few percent of
  QSOs.  Although the quality of our foreground QSO spectra is not as high as
  the SDSS data, absorption seen towards one of the background QSOs clearly
  does not show up at the same strength in the spectrum of the corresponding
  foreground QSO. This implies that the absorbing gas is distributed
  inhomogeneously around the QSO, presumably as a direct consequence of the
  anisotropic emission from the central AGN.  We discuss possible origins for
  the Mg~II lines, including: absorption by gas from the foreground QSO host
  galaxy; companion galaxies fuelling the QSO through gravitational
  interactions; and tidal debris left by galaxy mergers or interactions which
  initiated the QSO activity. No single explanation is entirely satisfactory,
  and we may well be seeing a mixture of phenomena.

\end{abstract}


\section{Introduction}

One of the few ways to study the distribution of gas around QSOs is through
absorption line studies. Until now, this has meant using the QSO itself as a
background source against which absorbing gas clouds might  be detected. But the origin
of the narrow-line ``associated'' $z_{\rm{abs}}\:\simeq\:z_{\rm{em}}$
systems is confusing, since even lines which have velocities $\approx
50,000$~\kms\ blueward of the emission redshift may be intrinsic to the QSO
\citep{richards99}, at least for high ionization species.  This might not be
surprising, considering that the sightline looks into the heart of the AGN,
but the ambiguity in translating absorption velocities into distances means we
can rarely be sure where any particular line comes from. It might arise in
material close to the AGN, gas ejected (and now distant) from the AGN, the
host galaxy, or a companion galaxy fuelling the black hole.

If, instead, we were able to find close pairs of QSOs on the sky with very
different redshifts, the background QSO could be used to probe the environment
of the foreground QSO.  Until recently, this experiment has been hard to do,
since few such QSO pairs were known.  However, the wide-area sky coverage and
spectroscopic follow-up capabilities of the {\it Sloan Digital Sky Survey}
\citep[SDSS;][]{york00} has resulted in the identification of copious numbers
of QSOs, making it possible to find chance alignments of quasars with very
different redshifts.

In this paper we describe the results of a study designed to search for
Mg~II~$\lambda\lambda 2796, 2803$~\AA\ absorption lines from foreground QSOs.
Mg~II is convenient because at $z\:\apg\: 0.4$ the UV doublet is redshifted
into the optical region covered by SDSS spectra, a redshift low enough to
investigate the environment of the foreground QSOs.  Although the sample of
pairs described herein is small, future studies will enable us to map the
gaseous structures around QSOs, as well as improve our understanding of how
quasars enrich the intergalactic medium (IGM)  they inhabit.

\begin{deluxetable*}{cclccrcccccc}
\tabletypesize{\footnotesize}
\tablecolumns{12}
\tablewidth{0pt} 
\tablecaption{QSO-QSO pairs \label{tab_pairs}}
\tablehead{
\colhead{}
& \colhead{}
& \colhead{}
& \colhead{}
& \colhead{}
& \colhead{}
& \colhead{} 
&& \colhead{} 
& \multicolumn{2}{c}{$W_r$ (Mg~II)\tablenotemark{a}}
& \colhead{} \\
\cline{10-11}
\colhead{Background QSO/}
& \colhead{}
& \colhead{}
& \colhead{$g$-band}
& \colhead{}
& \colhead{$\rho$}
& \colhead{$\rho$} 
&& \colhead{} 
& \colhead{$\lambda 2796$}
& \colhead{$\lambda 2803$}
& \colhead{\wAssoc\tablenotemark{b} } \\
\colhead{Foreground QSO}            
& \colhead{Plate$-$MJD$-$fibre}
& \colhead{$z_{\rm{em}}$}
& \colhead{psf mag}
& \colhead{$M_B$}
& \colhead{($''$)}
& \colhead{(\h) }
&& \colhead{$z_{\rm{abs}}$}
& \colhead{(\AA )}
& \colhead{(\AA )}
& \colhead{(\AA)} 
}
\startdata
SDSS J095454.99+373419.9   & 1596$-$52998$-$330 & 1.884  & 18.9  & \ldots  & 3.1    & 26     && 1.5496 & $1.10\pm0.17$ & $0.33\pm 0.19$ & \ldots \\
SDSS J095454.73+373419.7   & \ldots             & 1.544  & 19.6  & $-$24.8 & \ldots & \ldots && \ldots & \ldots        & \ldots         & $< 0.42$ \\

SDSS J083649.55+484154.0   & 0550$-$51959$-$426 & 1.711  & 18.5  & \ldots  & 4.1    & 29     && 0.6563 & $1.90\pm0.11$ & $1.24\pm 0.12$ & \ldots \\
SDSS J083649.45+484150.0   & \ldots             & 0.657  & 19.3  & $-$23.1 & \ldots & \ldots && \ldots & \ldots        & \ldots         & $< 0.56$ \\

SDSS J231253.03+144453.4   & 0743$-$52262$-$596 & 1.521  & 17.8  & \ldots  & 6.4    & 47     && 0.7672 & $0.39\pm0.12$ & $0.19\pm0.12$  & \ldots \\
SDSS J231252.79+144458.6   & \ldots             & 0.768  & 20.3  & $-$22.7 & \ldots & \ldots && \ldots & \ldots        &  \ldots        & $< 1.16$ \\

SDSS J211230.33$-$063332.1 & 0638$-$52081$-$551 & 1.544  & 19.5  & \ldots  & 15.2   & 98     && 0.5411 & $1.59\pm0.35$ & $1.69\pm0.35$  & \ldots \\
SDSS J211229.31$-$063331.4 & \ldots             & 0.551  & 20.2  & $-$22.1 & \ldots & \ldots && \ldots & \ldots        &  \ldots        & $< 1.60$ 

\enddata
\tablenotetext{a}{Rest equivalent widths of Mg~II absorption in the
  background (SDSS) QSO spectrum.}
\tablenotetext{b}{$2\sigma$ upper limit to the rest equivalent width of
  associated ($z_{\rm{abs}}\simeq z_{\rm{em}}$) absorption in the
  foreground (APO) QSO spectrum.}
\end{deluxetable*}

\section{QSO Pairs Selected and Data Analysis}

The QSO pairs listed in Table~1 were discovered as part of a search for binary
QSOs by \citet[][ hereafter H06]{hennawi06a}.  QSO pairs which are {\it not}
physically associated are listed in Table~9 of H06, and for our program we
selected pairs according to several criteria. First, the background QSO must
have been observed by SDSS; spectroscopic confirmation of many of H06's QSOs
were made using the Apache Point Observatory (APO) 3.5~m Astrophysical
Research Consortium (ARC) telescope, at a resolution and signal-to-noise (S/N)
too low to allow detection of Mg~II lines. Hence, APO spectra of {\it
  background} QSOs could not be used.  Second, the impact parameter between
the QSOs was chosen to be $< 150$~\h \footnote{\footnotesize $h_{70}\: =
  \:H_0/70$~\kms~Mpc$^{-1}$, $\Omega_m\:=\:0.3$ and
  $\Omega_\lambda\:=\:0.7$.}, a distance small enough to probe the inner
regions of the foreground QSO's environment.  Third, the foreground QSO had to
be between $0.4\: < \:z\:<\:1.6$.

The designations/positions and $g$-band magnitudes of the QSOs in Table~1 are
taken from {\it Data Release 4} of the SDSS Archive. The APO spectra of the
foreground QSOs were obtained between September 2003 and March 2004, using the
Double Imaging Spectrograph (DIS) configured with a 1.5 arcsec slit and 
low dispersion gratings, giving a resolution of 7$-$8~\AA\ FWHM (see
H06 for further details).  The redshifts of the foreground QSOs are taken from
Table~9 of H06, while their absolute $B$-band magnitudes come from the
cross-filter $K$-correction $K_{Bi}(z)$ between SDSS $i$-band and Johnson
$B$-band computed from the composite quasar spectrum of \citet{vandenberk_qso}
and Johnson-$B$ and SDSS-$i$ filter curves.  Background QSO spectra were
retrieved from the SDSS Archive.  At the wavelengths where Mg~II is expected
the resolution of the SDSS spectra is 2.2~\AA\ at 4500~\AA, and 3.6~\AA\ at
7000~\AA , FWHM, or $\sim 150$~\kms .

Mg~II lines were detected in the spectra of all four background QSOs.  The
SDSS spectra are plotted in Fig.~1, along with the lower resolution APO
spectra of the foreground quasars.  The redshift of the Mg~II absorption is
remarkably close to the systemic redshift of the QSO for three of the pairs.
Fig.~1 shows that the Mg~II emission line profile of \qsocfg\ appears offset
from the absorption, but the redshift of the QSO is actually derived from a
narrow [O~II]~$\lambda 3727$ line, which should give a more reliable systemic
redshift. The difference in velocity between absorption and emission for
\qsodfg\ is 1900~\kms , but the blue wing of the emission line appears quite
non-Gaussian, and may be affected by associated Mg~II absorption (see \S3).

To measure the properties of the absorption lines, we normalized the QSO
continua by defining a least-square fit of Legendre polynomials to the
continuum flux \citep{Semb92}. This results in a best-fit to the continuum,
along with upper and lower continua which represent the $\pm 1\sigma$ errors
in the adopted continuum.  Line wavelengths were measured by fitting Gaussian
profiles to the data. The final redshifts of the absorption systems were
determined by averaging the redshift of each line of the doublet.  Line
equivalent widths (EWs) were measured in the standard way: $W = \sum^{n}_i\:
(1-F_i) \:\delta \lambda$, where $F_i$ is the normalized flux at the i'th
pixel, $\delta\lambda$ is the wavelength dispersion, and the sum is calculated
over $n$ pixels. The variance from Poisson noise is given by $\sigma^2(W) =
\sum^{n}_i\: (\sigma_i\:\delta \lambda )^2$.  Error arrays from the SDSS
pipeline provided values of $\sigma_i$.  We took $n$ to be a value four times
the resolution, or nine pixels. Rest frame EWs ($W_r$) are given in Table~1.
Errors from continuum fitting were measured by taking the difference between
$W$ measured from the best-fit continuum and $W$ measured from the spectrum
normalized by the upper and lower $1\sigma$ error continua.  A final EW error
was calculated from the quadratic sum of the Poisson noise error and the
continuum error.

The ratio DR$=W_r(\lambda 2796)/W_r(\lambda 2803)$ towards \qsoabg\ is
$3.3\pm2.2$, which is consistent with the value of 2.0 for this doublet in
optically thin gas to within the $1\sigma$ errors; we also detect strong
C~IV~$\lambda\lambda 1548,1550 $ 
absorption at precisely the same redshift as the Mg~II line (Fig.~1), which
supports the reality of the detection. Towards \qsodbg , there appears to be
an additional absorption component redward of the Mg~II doublet. This may
simply be from Poisson noise, but the feature is significant at the $3\sigma$
level. It may therefore be more indicative of additional high velocity
absorption from the complex. Nevertheless, the identification of Mg~II at $
z=0.5411$ seems secure, since the two members of the doublet are detected at
a $4.5 \sigma$ significance, and the wavelengths of the two lines
give identical redshifts to within 20~\kms, or one-eighth of a resolution
element, assuming the lines are the two members of the Mg~II doublet.

We also calculated $2\sigma$ rest-frame EW limits, \wAssoc , for associated
absorption, i.e., absorption in the foreground QSO spectra at their emission
redshifts in the lower resolution APO spectra. Values of \wAssoc\ are given in
Table 1, though these may under-represent the true values; at the resolution
of the APO spectra, absorption lines may not be apparent at the peak of an
emission line, and the continuum placement may be under-estimated.

\begin{figure*}[t]
\vspace*{-5cm}\hspace*{0cm}\psfig{figure=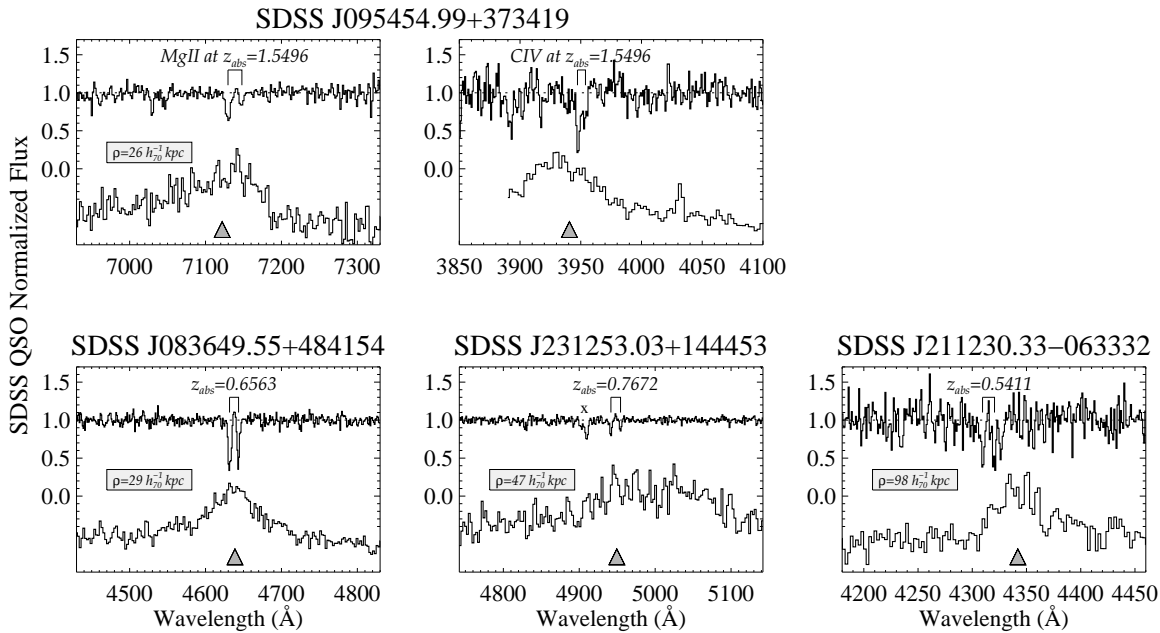,height=13cm,angle=90}
\caption{
Four panels showing Mg~II absorption in the normalized background (SDSS) QSO spectrum
(top of each panel), and one panel showing C~IV absorption towards \qsoabg .
Mg~II (and C~IV) emission in the foreground (APO) QSO spectrum is plotted 
at the bottom of each panel.
The foreground QSO spectra are not flux calibrated, and are scaled as needed.
Tick marks show the {\it predicted} position of Mg~II (and C~IV)
given the designated redshift.
The `X' marked in the spectrum of \qsocbg\ is an
Fe~II~$\lambda 2382$ line at $z=1.0604$. Triangles represent the expected
position of the narrow absorption lines in the background QSO, as well as the
Mg~II emission of the foreground QSO, based on the systemic redshift of the
foreground QSO.
 \label{fig_four}}
\end{figure*}

\section{Results}

Fig.~1 shows that Mg~II absorption is detected close to the redshift of each
of the four foreground QSOs. No other $z\:<\:1.6$ pairs for which the redshift
of the foreground QSO was confirmed spectroscopically were available, so there
are no examples of QSOs probed within 98~\h\ which failed to show Mg~II
absorption in background QSO spectra. This suggests that the Mg~II
cross-section around QSOs is high out to 98~\h, although the statistics are
clearly very poor considering the small number of pairs studied.  The Mg~II
absorbing galaxies found by \citet{stei95conf} extend only half as far,
although our results are more consistent with recent work by
\citet{church_china} and \citet{zibetti05}.  Given the paucity of Mg~II
systems along random QSO sightlines --- $< 1$ per unit redshift for lines with
the strengths listed in Table~1 \citep[e.g.][]{nestor05} --- the
absorption is unlikely to merely be a chance alignment in redshift space.

We also find that the absorption towards the background QSO \qsobbg\ --- that
is, in the direction {\it transverse} to the foreground QSO --- does not
appear at $z_{\rm{abs}}\:\simeq\:z_{\rm{em}}$ along the sightline to the
foreground quasar --- i.e. in the {\it radial} direction of the foreground
QSO.  The Mg~II absorption is strong ($W_r\:=\:2$~\AA ), and absorption of
similar strength would be detected in the APO spectrum of the foreground QSO,
which has a $2\sigma$ rest EW limit of \wAssoc$\:\approx 0.6$~\AA . For the
three other QSO pairs the lower resolution and S/N of the APO spectra make it
hard to detect absorption in the radial direction with the same strength as
that seen in the transverse direction.  The asymmetric emission line profile of the
foreground QSO \qsodfg\ could, however, be the result of intrinsic absorption,
and might explain the large velocity difference between absorption and
emission.

Our observations suggest that the incidence of Mg~II absorbing gas measured
transversely to a QSO is quite different from that measured radially towards
a QSO. The incidence of associated
($z_{\rm{abs}}\:\simeq\:z_{\rm{em}}$) narrow-line Mg~II absorption with
strengths similar to those detectable in SDSS data is $<11$~\%
\citep{SS92,aldcroft94}. In comparison, we find Mg~II in 4/4 systems in the
transverse direction, which appears to indicate that there exists a deficit of
Mg~II gas in the radial direction to QSOs relative to that seen in the
transverse direction. The difference is likely related to the
anisotropic radiation field of the QSO. Although the existence of Mg~II in a
gas cloud depends on many factors (strength of the ionizing field, gas
density, etc.)  the simplest, qualitative explanation for our results is that
the AGN radiation field strongly ionizes gas in the radial direction so that
Mg~II is not present, whereas gas situated in the transverse direction does
not see the same intense radiation, and is therefore less highly ionized.

\section{Discussion}

What are the possible origins for the Mg~II absorbing gas seen in the
transverse direction towards the background QSOs?  One possibility is that
Mg~II arises in host-galaxy disks; the three $z<1$ QSOs in Table~1 lie at the
faint end of the QSO luminosity function \citep{richards05}, and hosts of
low-luminosity, radio-quiet\footnote{\footnotesize None of the foreground QSOs
  are detected in the 1.4~GHz NRAO/VLA Sky Survey (NVSS) catalog
  \citep{condon98}; a 28~mJy source is detected in the field of \qsocbg , but
  it appears to be the background QSO.} QSOs often reside in disk galaxies
\citep{hamilton02}.  \qsobfg\ is probed at an impact parameter of 29~\h ; the
background QSO spectrum shows strong Fe~II~$\lambda 2382$
($W_r=1.53\pm0.13$~\AA) and Fe~II~$\lambda 2600$ ($1.01\pm0.11$~\AA ) lines.
The resulting Mg~II~($\lambda 2796$)/Fe~II~($\lambda 2600$) rest EW ratio of
$1.9\pm0.2$ means that there is a $\simeq 40$~\% probability that this
absorber is also a damped Lyman-$\alpha$ system, with $\log
N$(H~I)$\:\apg\:20$ \citep{rao06}. Even if $N$(H~I) is somewhat less
than this, the strengths of the absorption lines suggest that they might arise
in a galactic disk, since disks are indicative of regions of high column
densities.

Conversely, the absorption from \qsodfg\ is unlikely to arise in a host disk;
although the absorption is strong, few galaxy disks with radii of 98~\h\ are
known.  Of course, many QSOs are not hosted by disk galaxies at all.  QSOs
more luminous than $M_V < -24$ are typically found in massive elliptical
galaxies brighter than $L^*$ \citep[e.g.][ and refs.\ therein]{ociw_dunlop},
and one of our foreground QSOs, \qsoafg, is indeed more luminous than this.
\citet{floyd04} have shown that the average half-light radius
$\langle\:R_{1/2}\rangle$ is $\sim 14$~\h\ for the elliptical hosts of these
bright QSOs. The radius at which \qsoafg\ is probed, 26~\h , is only twice
$\langle R_{1/2}\rangle$, and is consistent with $R_{1/2}$ for some
of the larger ellipticals.  However, it is unclear whether strong Mg~II
absorption could be associated with a giant elliptical galaxy. Some
ellipticals contain small amounts of cold gas near their centers, but most of
the gas is expected to be too hot for Mg~II to survive.  There exists little
data on the absorbing properties of ellipticals. At low redshift, no Mg~II
absorption was seen in UV spectra of SN~1998S which exploded in the Fornax
elliptical NGC~1380 \citep{bowen_95}, although the lack of absorption may have
been due to a short path length through the galaxy to the supernova.  On the
other hand, \citet{steidel_mg2summ} found that {\it all} types of galaxies
were likely to be Mg~II absorbers.

A second possible explanation for the Mg~II absorption is that the lines arise
from a companion galaxy of the foreground QSO.  Low redshift QSOs reside in
galaxy groups of moderate richness \citep{nbahcall91} with the highest
overdensities occurring within $\sim 100$~kpc \citep{fisher96, serber06};
hence the probability of intercepting a galaxy in the same group as a QSO
is likely to be high.  Mg~II might arise from a companion galaxy directly interacting
with the quasar.  QSO activity probably begins when galaxies merge, and quasar
hosts are often found with close companions \citep{yee87,bahcall97}, even if
tidal remnants are not always apparent \citep{lim99}.  Interactions are
certainly an excellent method of distributing gas over a large area.
Absorption lines have been recorded from tidal debris
\citep{bowen_93j,deboer93,norman96}, and enrichment of the IGM by tidal
stripping has been invoked to explain absorption systems that have near solar
metallicities, but for which no absorbing galaxy can be found close to the
sightline \citep{phl1811_2,aracil06}.

Finally, we note that numerical simulations predict that
supermassive black holes
form when galaxies of similar mass merge \citep{kauffmann00, volonteri03,
  menci03}; 
again, the merger of two galaxies may produce tidal debris over a wide
area. In addition, inflows of gas onto a black hole may produce
intense star formation and the onset of powerful winds \citep[and refs.\ 
therein]{matteo05,hopkins05}. Intercepting such a wind might give rise to
absorption lines; it could also explain the large velocity difference between
emission and absorption in \qsodbg\ (if the systemic redshift of the
foreground QSO is correct). Whether such an explanation is tenable obviously
depends on the morphology of the wind and its orientation to the background
QSO sightline.

\section{Future Work}

Much work remains in order to characterize the Mg~II cross-section of QSOs.
Obviously, a sample of four QSO pairs is too small to enable us to draw firm
conclusions, and observations of more pairs are needed to test if the
$\sim\:100$\% covering factor is a robust estimate, and whether this changes
depending on the strength of the Mg~II lines selected.  Similarly, we need to
probe beyond 98~\h\ to find the limit of the Mg~II absorption.  With the
numerous QSOs detected by the SDSS, searching for Mg~II at larger radii is now
possible, and will be the subject of a future paper.  Deep, high resolution
imaging should help reveal the origin of the absorption if, e.g., companion
galaxies to the foreground QSOs, or evidence for previous mergers or
interactions, can be found.  Selecting foreground QSOs with $z\ll 1$ makes
detecting faint companions or asymmetric morphologies more feasible.
Echelle-resolution spectra of background QSOs will be important for finding
complex velocity structure in the absorbing gas, which may be related to its
origin in a wind or in tidal debris.

Spectra of foreground QSOs at resolutions and S/N comparable to SDSS spectra
will also be important for finding differences between absorption in the
radial and transverse directions of QSOs.  Since differences are likely
to be associated with changes in the ionization conditions of the gas,
observations of other higher ionization absorption lines in both background
and foreground QSO spectra will be pertinent.  As noted in \S2, the Mg~II
system towards the background QSO \qsoabg\ shows strong C~IV absorption,
although we have insufficient data to know whether C~IV absorption is also
seen towards the foreground QSO in the radial direction.  Such studies can be
carried out for high-$z$ foreground QSOs when the high ionization lines are
redshifted beyond the atmospheric cut-off.

\acknowledgments

D.V.B is funded through LTSA NASA grant NNG05GE26G. 
J.F.H is supported through NASA Hubble Fellowship grant
01172.01-A, awarded by the Space Telescope Science Institute.
B.M. acknowledges financial support from the F. Gould Foundation.
D.C. is supported by NASA through Chandra Postdoctoral Fellowship award
PF4-50033. M.A.S. is supported through NSF grant AST$-$0307409.
Funding for
the SDSS Archive has been provided by the Alfred P.~Sloan
Foundation, the Participating Institutions, NASA, the National Science
Foundation, the Dept.\ of Energy, the Japanese Monbukagakusho, and the Max
Planck Society.

\bibliographystyle{apj}
\bibliography{apj-jour,bib2}

\end{document}